\documentclass[reprint, superscriptaddress, amsmath, amssymb, aps, prl]{revtex4-2}

\usepackage{graphicx} 
\usepackage{dcolumn}  
\usepackage{bm}       
\usepackage{hyperref} 
\usepackage{float}
\usepackage{subfigure}
\usepackage{amsmath}
\usepackage{comment}
\usepackage{xcolor}

\begin{document}

\title{Quantum Destabilization of Skyrmions in Centrosymmetric Frustrated Magnets}

\author{Amit Kumar}
\affiliation{Theory Division, Saha Institute of Nuclear Physics,
A CI of Homi Bhabha National Institute, Kolkata 700064, India}

\author{Kalpataru Pradhan}
\affiliation{Theory Division, Saha Institute of Nuclear Physics,
A CI of Homi Bhabha National Institute, Kolkata 700064, India}

\date{\today}

\begin{abstract}
We investigate the role of the spin quantum number $s$ on the stability of skyrmions
in a $J_1-J_2-J_3$ centrosymmetric quantum Heisenberg model on a square lattice using
the neural network quantum states method. Our results reveal that the skyrmion
stability $Q$ is severely degraded when transitioning from the semiclassical regime
to the extreme quantum limit ($s=1/2$), where it ultimately vanishes. We demonstrate
that this destabilization is driven by quantum longitudinal fluctuations, with $Q$
exhibiting a power-law decay as a function of the reciprocal spin moment $1/s$.
Notably, the extreme quantum limit ($s=1/2$) deviates drastically from this scaling
behavior, exhibiting distinct physics compared to larger spin moments. Furthermore,
we reveal the microscopic  origin of this decay by establishing a quantitative
correspondence between skyrmion stability, entanglement, and local spin magnitude:
as the local second R\'enyi entropy (an indicator of entanglement) increases and
the local spin magnitude is suppressed, the skyrmion stability vanishes linearly.
This regime marks a quantum state where the skyrmion number $C$ remains as remanent
geometric feature of the spin orientations, yet the skyrmion stability $Q$ vanishes
due to the longitudinal suppression of the local spin magnitude. Our findings suggest
that classically robust skyrmion phases in frustrated lattices are fundamentally
restricted to high-spin materials, indicating that a spin moment must of at least
$s = 3/2$ is required for the realization of stable, atomic-scale topological textures.
\end{abstract}

\maketitle

\textit{Introduction}---The discovery of topological magnetic textures \cite{nagaosa2013topological, bogdanov2020physical, Dovzhenko2018, je2020direct}, 
particularly skyrmion crystals, has revolutionized our understanding of 
magnetism in both chiral and centrosymmetric materials. 
In chiral systems, the Dzyaloshinskii–Moriya interaction (DMI)
is mainly responsible for the formation of skyrmions\cite{dzyaloshinskii1958thermodynamic, Qin2018Stabilization}. 
The DMI energetically favors chiral spin textures and stabilizes skyrmions over a wide range of 
temperatures and magnetic fields\cite{Boni2020FieldTemp, Meisenheimer2024, cui2022anisotropic, Li2022AnisotropyOE, Carvalho2023Correlation}. 
In contrast, the existence of skyrmions in centrosymmetric magnets poses a conceptual challenge, 
as the absence of DMI removes the most common microscopic mechanism responsible for chiral spin textures. 
In such systems, skyrmions emerge from a delicate competition between exchange interactions \cite{PhysRevB.93.064430, Kawamura2025FrustrationinducedSC, PhysRevLett.118.247203, Paul2019RoleOH},
such as in the frustrated $J_1$–$J_2$–$J_3$ Heisenberg model \cite{Hayami2022RectangularAS, Leonov2015MultiplyPS, PhysRevX.9.041063}. 
This mechanism enables significantly smaller skyrmion textures \cite{Khanh_2020, Chakrabartty2021TunableRT},
with higher packing densities, positioning them as ideal candidates for 
next-generation, atomic-scale spintronics \cite{fert2017magnetic, marrows2021perspective, Wang2022JMMM}. 

Recent experimental progress has provided evidence that skyrmions can indeed exist in centrosymmetric 
materials such as Gd$_2$PdSi$_3$, GdRu$_2$Si$_2$, MnFe$_{0.2}$Co$_{0.8}$Ge etc.\cite{Paddison2024SpinDO, PhysRevB.110.134413, PhysRevLett.128.157206, PhysRevLett.129.137202, Chakrabartty2025Role}. 
Importantly, all of these materials carry a large spin moment (such as spin-$7/2$ for Gd). 
According to the existing literature, there are no known centrosymmetric materials with small spin moment, such as spin-$1/2$, that host skyrmions. 
In the classical regime, the stability of the skyrmion is measured through directional order, with the assumption that 
the local spin magnitude remains constant. However, as the spin moment $s$ is reduced toward the extreme quantum limit 
($s=1/2$), the local spin magnitude no longer remains a constant quantity, leaving the particle-like nature of the skyrmion vulnerable 
to quantum effects.

Despite significant research into chiral systems \cite{karube2020metastable,ham2021dmi, Mishra2023ApplPhysRev, Muhlbauer2009, Yu2010, Seki2012, PhysRevB.101.220405, Deng2020RoomtemperatureSP}, 
it remains unknown to date whether skyrmion stability 
truly persists or entirely vanishes for $s=1/2$ in centrosymmetric frustrated lattices. This leads to a 
fundamental question: Does the stability of skyrmion disappear abruptly at a specific 
threshold, or does it get destabilized continuously as a function of the spin moment? However, characterizing 
this transition is theoretically challenging, as standard perturbative frameworks fail to account for the 
non-linear longitudinal fluctuations inherent in the quantum limit. Consequently, there is a clear 
need for a non-perturbative approach that can link the skyrmion stability $Q$ to the microscopic 
suppression of the local spin magnitude that drives the destabilization.

In this paper, we employ neural network quantum states (NQS), a powerful non-perturbative variational framework, to investigate 
the quantum destabilization of skyrmions in centrosymmetric lattice systems. Unlike traditional expansions that assume small deviations from 
a classical background, the NQS approach enables us to capture the full non-linear response of the wavefunction, including the longitudinal 
fluctuations and entanglement that drive the destabilization transition. By training a variational ansatz to capture complex ground-state 
correlations, we simultaneously bypass the limitations of exact diagonalization and the sign problem of Quantum Monte Carlo. 
Using a $J_1$–$J_2$–$J_3$ quantum Heisenberg model, we show 
that as $s$ is reduced, the skyrmionic state undergoes a quantum destabilization. Specifically, we demonstrate that the 
proliferation of quantum longitudinal fluctuations, which drives the system toward a state of high entanglement, 
acts as a destabilizing factor for the magnetic texture. This process leads to a collapse of the skyrmion 
stability $Q$ despite a remanent skyrmion number $C$. Our scaling analysis reveals a 
power-law decay of $Q$ as a function of $1/s$ and 
identifies a sharp decrease in skyrmion stability near $s = 1$ suggesting that achieving a classically robust skyrmion phase requires a spin moment of at least $s = 3/2$.

\textit{Model Hamiltonian and Methods}---We consider the $J_1$--$J_2$--$J_3$ quantum Heisenberg model on a square lattice, restricted to the $x$-$y$ plane. 
The model includes an easy-axis Heisenberg anisotropy along the $z$ direction. 
The system is subjected to a strong magnetic field applied only to the boundary spins along the negative $z$ direction to fix their orientations and mimic a ferromagnetic embedding. 
A uniform bulk magnetic field is applied to the internal spins in the direction opposite to the boundary field.
The Hamiltonian of the system is given by:
\[
\begin{aligned}
H =\;& -J_1 \sum_{\langle ij \rangle} 
\left( S_i^x S_j^x + S_i^y S_j^y \right)
     -J_2 \sum_{\langle\langle ij \rangle\rangle} 
\left( S_i^x S_j^x + S_i^y S_j^y \right) \\
& -J_3 \sum_{\langle\langle\langle ij \rangle\rangle\rangle} 
\left( S_i^x S_j^x + S_i^y S_j^y \right)
     -A \sum_{\langle ij \rangle} S_i^z S_j^z  \\
& -B_z \sum_{i \in \mathrm{internal}} S_i^z
     +B \sum_{i \in \mathrm{boundary}} S_i^z
\end{aligned}
\]
where, $S_i^x$, $S_i^y$, and $S_i^z$ are the spin operators representing the corresponding spin components at lattice site $i$. 
The parameters $J_1$, $J_2$, and $J_3$ denote the nearest-neighbor, next-nearest-neighbor, and third-nearest-neighbor exchange interaction strengths, respectively. 
$A$ represents the strength of the Heisenberg anisotropy along the $z$ direction. 
$B_z$ denotes the strength of the uniform magnetic field applied to the internal bulk spins. 
Finally, $B$ denotes the strength of the external magnetic field applied only to the boundary spins in the opposite direction to that of bulk magnetic field.
In our calculations, we set $J_1 = 1$ and express all other parameters in units of $J_1$. 
We use $B = 10$, $7\times7$ lattice size, 
and average all the results over ten independent realizations, unless otherwise specified.

To determine the ground state wavefunction of the Hamiltonian, we employ the NQS method following Refs.~\cite{Joshi2023GroundSP, Joshi2024QuantumSD, kumar2025impactrandombonddisorder, Szab__2020, excited_state}. 
The variational wavefunction is given by $\sum_{\boldsymbol{\sigma}}\psi_\theta(\boldsymbol{\sigma})|\boldsymbol{\sigma}\rangle$, 
where $|\boldsymbol{\sigma}\rangle$ are computational basis states, chosen here as the eigenstates of the $S^{z}$ operators and $\theta$ denotes the set of trainable parameters. 
The logarithm of the complex wavefunction is 
modeled as $\ln \psi_\theta(\boldsymbol{\sigma})=\rho(\boldsymbol{\sigma})+i\phi(\boldsymbol{\sigma})$, 
where amplitude $\rho(\boldsymbol{\sigma})$ and phase $\phi(\boldsymbol{\sigma})$ are represented using two independent fully connected feed-forward neural networks with ReLU activation.
The network parameters are optimized by minimizing the energy expectation value $\langle\psi_\theta|\hat{H}|\psi_\theta\rangle$ using the 
Adam optimizer. Spin configurations are sampled using Markov Chain Monte Carlo, and the numerical implementation is carried out with the NetKet library \cite{netket3:2022, netket2:2019}. 
For comparison with the classical limit, simulations are performed using classical Monte Carlo methods with the Metropolis–Hastings algorithm.
Further details of the implementation are provided in the Appendix.

\textit{Topological Stability}---To investigate how the value of the local spin moment affects the stability of skyrmions, 
we first determine the parameter regime in which a single skyrmion can be realized. 
In centrosymmetric lattices, 
magnetic frustration provides a possible route to stabilizing a single skyrmion 
\cite{PhysRevB.93.064430, Kawamura2025FrustrationinducedSC}. 
This requires that the next-nearest-neighbor and third-nearest-neighbor exchange couplings, $J_2$ and $J_3$, 
to have signs opposite to that of the nearest-neighbor coupling $J_1$, i.e., they must be antiferromagnetic. 

To map the parameter space that supports a single skyrmion state,
we first define suitable indicators to detect and characterize the topological texture. 
Using the local spin expectation values 
$\langle \vec{S}_i \rangle$, we compute two related quantities: the skyrmion number $C$ 
and the skyrmion stability $Q$, following Refs.~\cite{Siegl_2022, Sotnikov_2021}. 
Both $C$ and $Q$ are evaluated using the same geometric expression constructed from the scalar triple product 
of spins defined on a triangular plaquettes of a triangulated square lattice:
\[
C, Q = \frac{1}{2\pi} \left| \sum_{\Delta} \tan^{-1} \left( 
\frac{ \vec{n}_i \cdot (\vec{n}_j \times \vec{n}_k) }
{ 1 + \vec{n}_i \cdot \vec{n}_j + \vec{n}_j \cdot \vec{n}_k + \vec{n}_k \cdot \vec{n}_i } 
\right) \right|
\]
where the summation runs over all elementary triangles $\Delta$ of the lattice. 
The indices $i$, $j$, and $k$ denote the three lattice sites forming the vertices of each triangle.
The difference between $C$ and $Q$ arises from the definition of the local spin vector $\vec{n}_i$, 
which is constructed from the expectation value of the spin operator at site $i$, 
$\langle \vec{S}_i \rangle = (\langle S_i^x \rangle, \langle S_i^y \rangle, \langle S_i^z \rangle)^T$. 
For the calculation of $C$, we defined the normalized vector as  
$\vec{n}_i = \langle \vec{S}_i \rangle / |\langle \vec{S}_i \rangle|$,
while for calculation of $Q$ we use $\vec{n}_i = \langle \vec{S}_i \rangle / s$, 
where $|\langle \vec{S}_i \rangle| = \sqrt{ \langle S_i^x \rangle^2 + \langle S_i^y \rangle^2 + \langle S_i^z \rangle^2 }$
and $s$ is the spin quantum number.
Note that the skyrmion number $C$, depends only on the angular winding and topological geometry of 
the spin texture irrespective of the value of local spin magnitude, thereby results in strictly integer values. 
In contrast, the skyrmion stability $Q$ is computed using the unnormalized spin expectation value scaled 
by the total spin magnitude. Because quantum fluctuations can suppress the local spin magnitude below its 
maximum classical value $s$, $Q$ explicitly accounts for changes in the local spin length. 
Accordingly, $Q$ can take non-integer values ranging from 0 to 1, where 1 corresponds to a perfectly stable 
skyrmion configuration and 0 indicates complete destabilization of the spin texture.

\begin{figure}[htb]
    \centering
    \includegraphics[width=1.0\linewidth]{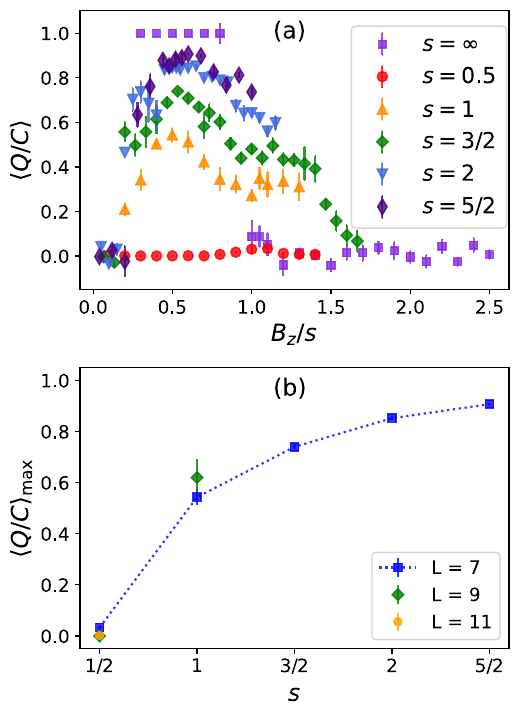}
    \caption{Role of spin quantum number on skyrmion stability:
      (a) Ratio of skyrmion stability to skyrmion number $\langle Q/C \rangle$ as a function 
    of the normalized bulk magnetic field $B_z/s$, averaged over ten independent realizations.
    Note that the normalization is not applied to $B_z$ for classical case ($s=\infty$). 
    (b) Extracted maximum value of $\langle Q/C \rangle$ from (a)
    as a function of spin moment $s$ (additional green and orange data points represent lattice size $L=9$ and $11$, respectively).
    The sharp decline toward the $s=1/2$ limit indicates the strong impact of quantum longitudinal fluctuations 
    on skyrmion stability, even while the skyrmion number $C$ remains preserved.}
    \label{fig:QvsBz}
\end{figure}

By analyzing 
the spin texture together with the corresponding 
value of $C$, 
we can determine whether a skyrmion is present. 
Specifically, if we obtain $C = 1$, we identify the state as a single skyrmion. 
Since the absence of DMI removes the chiral mechanism that uniquely selects a specific helicity, centrosymmetric systems can 
support all different types of skyrmions such as Bloch, Néel, and antiskyrmions within the same parameter regime.
To account for this, we consider the absolute values of $|C|$ and $|Q|$ to treat all the configurations on an equal footing.

In addition, because skyrmions in centrosymmetric models arise from a delicate balance among competing exchange interactions 
and are highly sensitive to external perturbations \cite{Khanh_2020, Chakrabartty2021TunableRT}, 
the parameters must be tuned carefully to stabilize the skyrmion. 
For $s=1/2$ we find negligible value of skyrmion stability ($Q < 0.1$) across variation of all the parameters ($J_2, J_3, A, B_z$), 
even when the skyrmion number $C$ remains finite.

For higher finite spin moments ($s = 1, 3/2, 2, 5/2$), we perform quantum NQS simulations, 
whereas in the classical limit ($s=\infty$), simulations are carried out using classical Monte Carlo.
From systematic analysis of the parameter space, we find that choosing 
$J_2 = -0.425$, $J_3 = 0.5 J_2$, $A = 0.4$, 
and varying the bulk magnetic field $B_z$, 
stabilizes a skyrmion state for all higher spin moments.
Since skyrmions in centrosymmetric lattices are typically smaller in size~\cite{Khanh_2020, Chakrabartty2021TunableRT}, 
for certain values of the bulk magnetic field $B_z$ we obtain $C = 2$. 
We account for these multi-skyrmion states by evaluating the ratio of skyrmion stability $Q$ to skyrmion number $C$. 
To represent the variation of skyrmion stability with $B_z$, we select ten independent realizations for which a finite value 
of $C$ is obtained and compute the average $Q/C$ ratio. 
Finally to facilitate a direct comparison of the skyrmion stability across different spin moments and the classical limit, 
we scale the bulk magnetic field as $B_z/s$ for all finite spins, while no scaling is applied in the classical limit. 

We begin our analysis by plotting the normalized skyrmion stability $\langle Q/C \rangle$ as a 
function of the normalized bulk magnetic field $B_z/s$ for different spin moments $s$, 
ranging from the extreme quantum limit ($s = 1/2$) to the semiclassical regime ($s \geq 3/2$) as shown in Fig.~\ref{fig:QvsBz}(a). 
In the semiclassical regime at $s=2$ and $s=5/2$, we observe a robust topological plateau where $\langle Q/C \rangle \approx 0.9$, 
indicating that the skyrmion stability is nearly perfectly aligned with the skyrmion number $C$. However, as $s$ decreases, 
the stability plateau is progressively suppressed and narrowed.
At the extreme quantum limit ($s = 1/2$), the stability almost vanishes entirely across all field strengths. 

The magnitude of 
this collapse is summarized in Fig.~\ref{fig:QvsBz}(b), which plots the extracted maximum normalized skyrmion stability $\langle Q/C \rangle _\mathrm  {{max}}$
as a function of the spin moment $s$. The data 
reveal a sharp, non-linear decline in stability as the system approaches the quantum limit. This abrupt drop, despite the 
persistent skyrmion number $C$ in the underlying texture, signals a transition from a stable skyrmion to 
a strong quantum regime, where the skyrmion loses its physical stability. 
To ensure that this destabilization is not a finite-size artifact, we verified the trend on larger lattices, 
confirming that the suppression of stability remains robust at $s=1/2$ for $9\times9$ and $11\times11$ sizes, as well as for $s=1$ on a $9\times9$ lattice.

\begin{figure}[htb]
  \centering
  \includegraphics[width=1.0\linewidth]{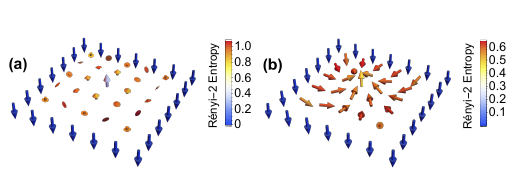}
  \caption{Real-space spin texture and local entanglement for different spin moments: 
  Panels show the spin texture for (a) $s = 1/2$ and (b) $s = 3/2$, obtained at 
  the maximum normalized skyrmion stability $\langle Q/C \rangle_\mathrm{{max}}$. The color of the 
  arrows represents the local second Rényi entropy, while the arrow length indicates the magnitude of 
  the local spin expectation value $|\langle \vec{S}_i \rangle|$. Comparing the two limits reveals that while the $s=1/2$
  texture (a) retains the topological directional winding, its local spin magnitudes are dramatically suppressed and 
  accompanied by higher entanglement entropy compared to the semiclassical $s=3/2$ case in (b).}
  \label{fig:spin_texture}
\end{figure}

\textit{Microscopic Destabilizers}---To analyze what happens to the skyrmion at low spin moments and why it becomes highly unstable, 
we study both the real-space spin texture and the corresponding quantum entanglement properties. 
To quantify the entanglement, we use the second Rényi entropy $S_2$, 
evaluated from the expectation value of a swap operator. 
To compute $S_2$ per lattice site, we partition the lattice into two subsystems: 
region $A$, consisting of a single spin, and region $B$, containing the remaining spins of the system.
The second Rényi entropy is defined as \cite{PhysRevLett.104.157201}:
\[
S_2(\rho_A) = -\log_2 \left( \text{Tr}(\rho_A^2) \right) 
= -\log_2 \left( \langle \text{Swap}_A \rangle \right)
\]
where $\langle \text{Swap}_A \rangle$ denotes the expectation value of the swap operator acting on subsystem $A$. 
In practice, we estimate $\langle \text{Swap}_A \rangle$ using Monte Carlo sampling over $N_s$ pairs 
of independent spin configurations \cite{PhysRevResearch.2.023358}, given by:
\[
\langle \text{Swap}_A \rangle 
\approx \frac{1}{N_s} \sum_{i=1}^{N_s} 
\frac{
\psi_\theta(\tilde{\boldsymbol{\sigma}}^{(i)}_A, \boldsymbol{\sigma}^{(i)}_B) \,
\psi_\theta(\boldsymbol{\sigma}^{(i)}_A, \tilde{\boldsymbol{\sigma}}^{(i)}_B)
}{
\psi_\theta(\boldsymbol{\sigma}^{(i)}_A, \boldsymbol{\sigma}^{(i)}_B) \,
\psi_\theta(\tilde{\boldsymbol{\sigma}}^{(i)}_A, \tilde{\boldsymbol{\sigma}}^{(i)}_B)
}
\]
where, $\boldsymbol{\sigma}^{(i)}$ and $\tilde{\boldsymbol{\sigma}}^{(i)}$ 
are spin configurations sampled independently from two copies of the variational wavefunction.

We visualize the real-space 
spin texture and local entanglement profiles for $s = 1/2$ and $s = 3/2$ in Fig.~\ref{fig:spin_texture}. In the semiclassical $s = 3/2$ case [Fig. 2(b)], 
the skyrmion appears as a robust magnetic structure. The local spin magnitudes $|\langle \vec{S}_i \rangle|$ 
remain close to their maximum possible value, while the microscopic entanglement (indicated by the color scale) is heterogeneously distributed, 
diminishing toward the core. This suggests that the topological stability is maintained by the central region of the skyrmion
with the quantum longitudinal fluctuations largely localized in the intermediate region between the core and the boundary spins.

In contrast, for $s = 1/2$ [Fig. 2(a)], the texture undergoes a dramatic visual transformation. While the arrows still define the 
geometric winding (resulting in $C = 1$), indicating that the geometric structure of the skyrmion persists, 
their physical lengths are visibly quenched across the lattice. This local spin magnitude suppression provides the first visual 
evidence of destabilization: the skyrmion number $C$ survives while
local spin magnitudes vanish, leading to the diminished skyrmion stability $Q$ identified in Fig.~\ref{fig:QvsBz}. 
These qualitative observations demand a deeper quantitative analysis of the relationship between local spin magnitude, 
entanglement, and stability in the following sections.

\begin{figure}[htb]
    \centering
    \includegraphics[width=1.0\linewidth]{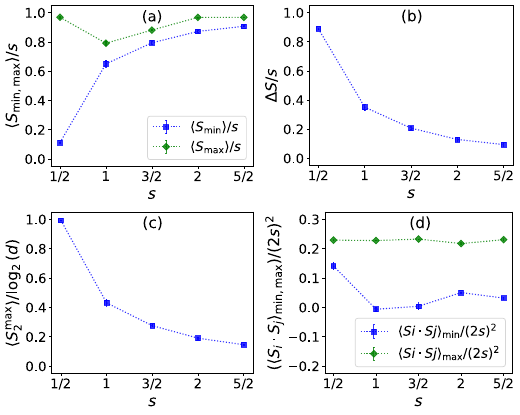}
    \caption{Evolution of local quantum observables as a function of spin quantum number $s$: 
    (a) Normalized minimum and maximum local spin magnitudes $\langle S_{\mathrm{min}}\rangle/s$, $\langle S_{\mathrm{max}}\rangle/s$, highlighting the contraction of the longitudinal magnetic moments. 
    (b) Normalized amplitude of quantum longitudinal fluctuations $\Delta S = 1 - S_\mathrm{min}$. 
    (c) Normalized maximum second Rényi entropy $S_2^\mathrm{max}/\log_2(d)$, quantifying the growth of local quantum entanglement. 
      (d) Normalized nearest-neighbor minimum and maximum spin–spin correlations $\langle Si \cdot Sj \rangle_{min}/(2s)^2$, $\langle Si \cdot Sj \rangle_{max}/(2s)^2$.
      The fact that these correlations do not approach the theoretical lower bound rules out local singlet formation as the
      primary mechanism behind skyrmion destabilization.}
    \label{fig:spinRenyi}
\end{figure}

To quantify the skyrmion destabilization observed in the real-space spin textures, 
we systematically analyze the local spin magnitudes and the second Rényi entropy, excluding boundary spins, across the different spin moments in Fig.~\ref{fig:spinRenyi}.
In Fig.~\ref{fig:spinRenyi}(a), we plot the extracted normalized minimum 
local spin magnitude $\langle S_{\mathrm{min}}\rangle/s$. As we move from the semiclassical regime toward the $s = 1/2$ extreme quantum limit, 
we observe a rapid drop in the magnitude of these weakest spins, confirming that quantum longitudinal fluctuations are most 
severe at specific sites within the texture. This suppression is accompanied by a sharp increase in the amplitude of the quantum longitudinal 
fluctuations $\Delta S / s = 1 - \langle S_{\mathrm{min}}\rangle/s$ as shown in Fig.~\ref{fig:spinRenyi}(b). 
At the same time the normalized maximum local spin magnitude $\langle S_{\mathrm{max}}\rangle/s$ in Fig.~\ref{fig:spinRenyi}(a) remains more or less same across all the $s$ values, 
suggesting that the local spin magnitudes 
do not just shrink throughout the lattice, but become increasingly non-uniform across the lattice. 

These results reveal that the 
$s = 1/2$ regime exhibits extreme longitudinal non-uniformity across the lattice, contrasting with the 
uniform magnitudes characteristic of the semi-classical limit.
Crucially, the growth of this non-uniformity is intrinsically linked to the rise of quantum entanglement. In Fig.~\ref{fig:spinRenyi}(c), we 
observe that the extracted maximum normalized second Rényi entropy $S_2^{\mathrm{max}}/\log_2(d)$, (where $d = 2s+1$ is the local Hilbert space dimension) increases significantly as $s$ approaches $1/2$. 
This trend demonstrates that the reduction of the local spin magnitude is a pure quantum effect, where the loss 
of local spin moment directly corresponds to an increase in quantum entanglement.

Next, to distinguish between many-body quantum effects and simple pairwise physics, we analyze the nearest-neighbor spin-spin correlations 
$\langle S_i \cdot S_j \rangle = \langle S_i^x S_j^x \rangle + \langle S_i^y S_j^y \rangle + \langle S_i^z S_j^z \rangle$, 
excluding boundary spins in Fig.~\ref{fig:spinRenyi}(d). The extracted normalized maximum correlation $\langle S_i \cdot S_j \rangle_{\mathrm{max}}/(2s)^2$ 
remains nearly constant across different spin moments, 
with a value slightly less than the classical saturation value ($0.25$), reflecting the canting experienced by spins in forming the skyrmion texture.
Furthermore, the extracted normalized minimum correlation $\langle S_i \cdot S_j \rangle_{\mathrm{min}}/(2s)^2$ remains close to zero, 
exhibiting only a slight increase as the system approaches $s=1/2$.

Notably, these values do not approach the theoretical lower bound of -0.75, which would be characteristic of singlet formation.
This behavior 
suggests that the observed entanglement and subsequent skyrmion destabilization are not driven 
by the formation of local singlets, a mechanism known to contribute to the destabilization of skyrmions in systems with random bond disorder~\cite{kumar2025impactrandombonddisorder}, 
but rather by a spatial suppression of local spin magnitudes induced by quantum longitudinal fluctuations. Together, 
these results demonstrate that the quantum skyrmion exists in a state of high longitudinal fluctuations and intense 
entanglement, providing the microscopic basis for the stability collapse observed in the quantum limit.

\begin{figure}[htb]
    \centering
    \includegraphics[width=1.0\linewidth]{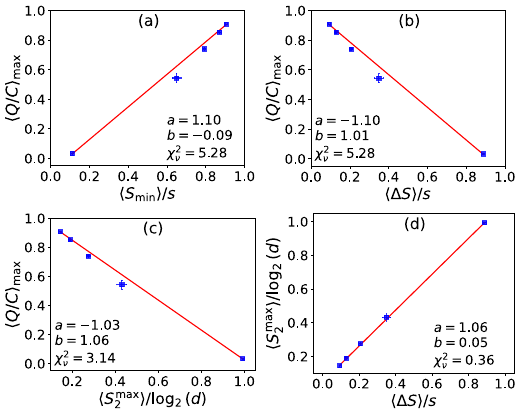}
    \caption{Direct correlations between the microscopic drivers of destabilization and macroscopically observed skyrmion stability: 
    Linear fits are performed across all panels using the relation $y = ax + b$, where $a$ and $b$ are fitting parameters.
    (a) Linear fit between maximum normalized skyrmion stability $\langle Q/C \rangle_{\mathrm{max}}$ and normalized minimum local spin magnitude 
    $\langle S_{\mathrm{min}}\rangle/s$ for different spin moment $s$.
    The strong positive linear correlation confirms that the structural destabilization is driven by the suppression of the local spin magnitude.
    (b) Linear fit between $\langle Q/C \rangle_{\mathrm{max}}$ and normalized amplitude of quantum longitudinal fluctuations $\Delta S / s = 1 - \langle S_{\mathrm{min}}\rangle/s$
    demonstrating that emergence of non-uniform local spin magnitudes directly reduces skyrmion stability.
    (c) Linear fit between $\langle Q/C \rangle_{\mathrm{max}}$ and normalized maximum second Rényi entropy $S_{2}^{\mathrm{max}}/\log_2(d)$
    signaling that the growth of many-body quantum entanglement degrades skyrmion stability.    
    (d) Linear fit between $S_{2}^{\mathrm{max}}/\log_2(d)$ and $\Delta S / s$ 
    revealing that intense quantum entanglement facilitates the spatial suppression and non-uniformity of the spin magnitudes.
    }
    \label{fig:linearfit}
\end{figure}

To explicitly link the local microscopic destabilizers established in Fig.~\ref{fig:spinRenyi} to the macroscopically observed skyrmion stability, 
we examine their direct correlations in Fig.~\ref{fig:linearfit}.
We perform a linear fit between the quantities shown on vertical axis $y$ and the horizontal axis $x$ using the equation $y = ax + b$, where $a$ and $b$ are fitting parameters. 
The corresponding reduced chi-square $\chi^2_\nu = \chi^2/\mathrm{dof}$ is calculated by 
accounting for uncertainties in both variables, where the effective variance is given by
$\sigma^2 = \sigma_{y}^2 + (\partial y/\partial x)^2 \sigma_{x}^2$, 
and $\chi^2 = \sum (y - y_{\mathrm{pred}})^2/\sigma^2$.
In Fig.~\ref{fig:linearfit}(a), we establish that the maximum normalized skyrmion stability $\langle Q/C \rangle_{\mathrm{max}}$ 
exhibits a strict linear dependence on the normalized minimum local spin magnitude $\langle S_{\mathrm{min}}\rangle/s$. 
This directly confirms that the physical stability of the skyrmion is limited by the value of its local magnetic moments. 
In the extreme quantum limit ($s=1/2$), 
the skyrmion is merely a miniaturized version of its classical counterpart but due to the severe suppression of its local spin magnitudes,
these moments can be easily flipped by very small perturbations~\cite{Siegl_2022}, leading to a unstabilized skyrmion.

This structural disruption is further quantified in Fig.~\ref{fig:linearfit}(b), 
where the $\langle Q/C \rangle_{\mathrm{max}}$ is fitted against the amplitude of quantum longitudinal fluctuations $\Delta S/s = 1 - \langle S_{\mathrm{min}}\rangle/s$. 
This trend confirms that the destabilization transition is  driven by the emergence of spatial non-uniformity in the local spin magnitudes across the texture,
corroborating our finding from Fig.~\ref{fig:spinRenyi}(a) that the normalized maximum local spin magnitude remains nearly constant across all $s$ values.
Furthermore, this non uniform suppression of local spin magnitude almost reaches its maximum limit (i.e. $\Delta S/s \approx 1$) for $s=1/2$, making the texture totally unstable.
The quantum origin of this instability is highlighted in Fig.~\ref{fig:linearfit}(c), 
which reveals that the skyrmion stability is inversely proportional to the maximum normalized second Rényi entropy. 
The negative linear slope identifies entanglement as a primary destabilizing factor in the extreme quantum limit. 
Finally, the microscopic sequence of this transition is presented in Fig.~\ref{fig:linearfit}(d), 
which shows the mutual evolution of the amplitude of quantum longitudinal fluctuations and entanglement with decreasing spin moment. 
This correlation suggests that 
the spatial non-uniformity of the local moments is intrinsically tied to the entanglement; specifically, 
higher entanglement is necessarily accompanied by larger suppression and pronounced spatial non-uniformity of the local spin magnitudes.

\begin{figure}[htb]
    \centering
    \includegraphics[width=1.0\linewidth]{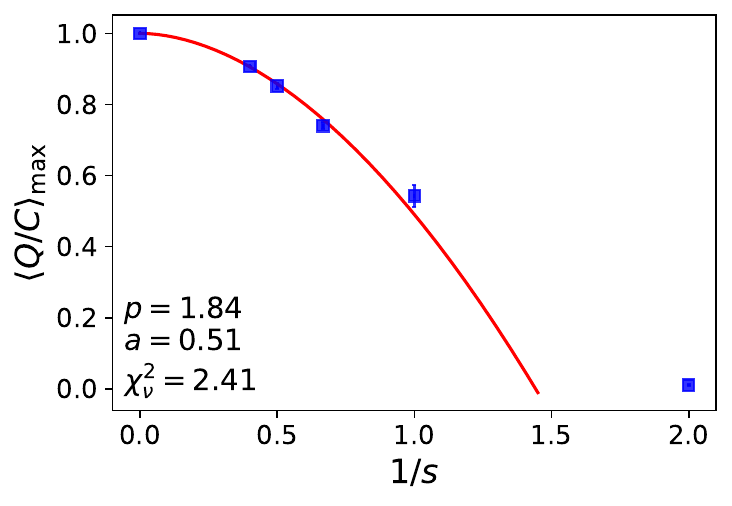}
    \caption{Scaling of the skyrmion destabilization transition:
      Maximum normalized skyrmion stability $\langle Q/C \rangle_{\mathrm{max}}$ as a function of 
    the reciprocal spin moment $1/s$. The red line represents a power-law fit 
    $\langle Q/C \rangle_{\mathrm{max}} = 1 - a|1/s|^p$, evaluated over the range $s \geq 1$. 
    The data show that while high-spin systems ($s \geq 2$) maintain high skyrmion stability, 
    the stability collapses as $s$ approaches the quantum limit. The $s=1/2$ data point (rightmost) is 
    excluded from the fit as it represents a fully destabilized phase where the skyrmion stability $Q$ is 
    near zero while the skyrmion number $C$ remains well-defined.}
    \label{fig:powerlawfit}
\end{figure}

\textit{Stability and Scaling Behavior}---The cumulative effect of these microscopic drivers results in the qualitative scaling behavior captured in Fig.~\ref{fig:powerlawfit}. We plot the 
maximum normalized skyrmion stability $\langle Q/C \rangle_{\mathrm{max}}$ as a function of the reciprocal spin moment $1/s$, 
with the classical boundary condition ($\langle Q/C \rangle_{\mathrm{max}} = 1$ at $1/s = 0$). For the range $s \geq 1$,
the data follow a near-quadratic power-law decay, calculated using the form $\langle Q/C \rangle_{\mathrm{max}} = 1 - a|1/s|^p$, with an exponent of $p \approx 1.8$.  
This suggests that stability is restored with increasing spin moment $s$, far more rapidly than 
the $O(1/s)$ reduction typically predicted by linear spin-wave theory~\cite{PhysRevResearch.2.043243, UtermohlenSpinWave}. 
In the semiclassical regime at $s = 2$ (i.e., $1/s = 0.5$), the system retains over 85\% of its stability, 
indicating that the skyrmion state is well-supported by the underlying local spin magnitude. 
As the spin moment increases to $s=5/2$ (1/s=0.4), the stability reaches approximately 90\%. 
On the other hand, as the system approaches the quantum regime at $s=1$ there is a rapid drop in stability, with the system retaining only 50\% of its classical value.
Notably, the $s=1/2$ data point is excluded from this fit, as it represents an extreme quantum limit where skyrmion stability is entirely suppressed.

The microscopic evidence of skyrmion destabilization presented in this work reveals a fundamental trade-off between skyrmion size and 
its stability. While $s=1/2$ chiral magnets remain a popular platform for topological spintronics due to DMI-stabilization, 
their relatively large skyrmion size limits high-density device miniaturization. Conversely, centrosymmetric frustrated magnets offer a route 
to atomic-scale skyrmions, however as our results demonstrate, these sub-nanometer skyrmions are highly susceptible to the destructive quantum 
longitudinal fluctuations in the low-spin limit. To prevent this destabilization while maintaining a sub-nanometer size, a shift in material focus is required.  
We conclude that the optimal candidates for next-generation, high-density topological devices are high-spin frustrated magnets with spin moment of at least $s = 3/2$, 
such as compounds utilizing $Fe^{3+}$ or $Mn^{2+}$ ions. These systems provide the ideal compromise: the large spin moment 
suppresses the destructive fluctuations identified in our study, while the exchange-frustrated mechanism retains the 
nanometer-scale size essential for future technologies.

\textit{Conclusion}---In summary, we have utilized NQS to uncover the microscopic mechanism governing the stability of skyrmions 
in centrosymmetric frustrated magnets. Our results demonstrate that the observed skyrmion  destabilization is 
driven by the mutual evolution of entanglement and spin magnitudes across the spin moments. In the extreme quantum limit ($s=1/2$), 
we showed that while the skyrmion number $C$ remains intact, the skyrmion stability $Q$ is virtually nullified as 
the local spin magnitudes are suppressed and become highly non-uniform. This destabilization occurs because $Q$ is undermined by 
intense many-body entanglement and severe quantum longitudinal fluctuations. 
Consequently, the skyrmion destabilization follows a near-quadratic power-law scaling behavior with respect to the reciprocal spin moment.

In addition, our findings provide a predictive roadmap for the material design of atomic-scale topological textures.
We establish that spin moments have to be at least $s = 3/2$ to suppress quantum longitudinal fluctuations and maintain a stable skyrmion. 
Hence, our results suggest that searching for stable, atomic-scale skyrmions in $s=1/2$ or $s=1$ frustrated systems may prove 
futile for practical applications 
due to their inherent lack of physical stability. Instead, high-spin frustrated magnets, 
such as those based on $Fe^{3+}$, $Mn^{2+}$ or $Gd^{3+}$ ions, 
emerge as the optimal candidates, successfully bridging the gap between sub-nanometer size 
and the robust topological protection required for next-generation spintronics.

\textit{Acknowledgement}---We acknowledge National Supercomputing Mission
(NSM) for providing computing resources of `PARAM
RUDRA' at S.N. Bose National Centre for Basic Sciences, which is implemented by C-DAC and supported
by the Ministry of Electronics and Information Technology (MeitY) and Department of Science and Technology
(DST), Government of India, and the use of CCS3 computing cluster at SINP.

\section{Appendix}
\textit{NQS Architecture and Training Details}---To determine the ground state wavefunction of the Hamiltonian, we employ the NQS method following Refs.~\cite{Joshi2023GroundSP, Joshi2023GroundSP, kumar2025impactrandombonddisorder}. 
The variational wavefunction is given by:
\[
|\psi_\theta\rangle = \sum_{\boldsymbol{\sigma}} \psi_\theta(\boldsymbol{\sigma}) \, |\boldsymbol{\sigma}\rangle,
\]
where $|\boldsymbol{\sigma}\rangle$ denotes the computational basis states, chosen here as the eigenstates of the $S^{z}$ operator. 
The symbol $\theta$ represents the complete set of trainable parameters of the network.
The logarithm of the complex wavefunction 
is modelled using two independent fully connected feed-forward neural networks and given by:
\[
\ln \psi_\theta(\boldsymbol{\sigma}) = \rho(\boldsymbol{\sigma}) + i \phi(\boldsymbol{\sigma}),
\]
where $\rho(\boldsymbol{\sigma})$ and $\phi(\boldsymbol{\sigma})$ correspond to the outputs of the 
networks describing the amplitude and the phase, respectively.

\begin{figure}[htb]
    \centering
    \includegraphics[width=0.7\linewidth]{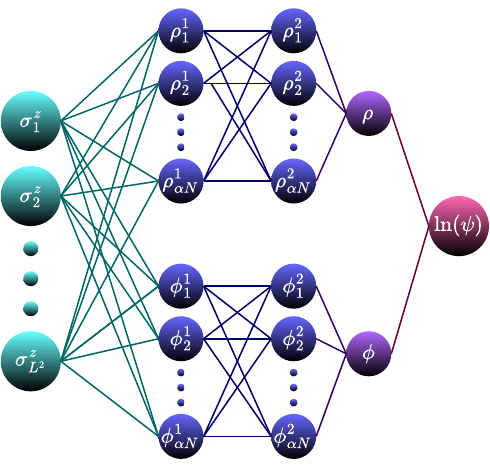}
    \caption{Architecture of the NQS: The input is a lattice spin configuration in the $S^z$ basis. Two separate neural networks are used to model the amplitude $\rho$ and 
    phase $\phi$. Each hidden layer contains \(\alpha N\) neurons, where $N$ is the total number of lattice sites. The output is the logarithm of the wavefunction.}
    \label{fig:nqs}
\end{figure}

The input to both networks is the spin configuration of the lattice. 
Each network (for $\rho$ and $\phi$) consists of two fully connected hidden layers.
Each hidden layer contains \(\alpha N\) neurons, where $N$ is the total number of lattice sites (see Fig.~\ref{fig:nqs}). 
Since dimension of Hilbert space increases rapidly with the spin moment $s$, we set $\alpha = 2$ for $s = 1/2, 1$; $\alpha = 3$ for $s = 3/2, 2$; and $\alpha = 4$ for $s = 5/2$. 
The ReLU (Rectified Linear Unit) activation function is used for all hidden layers.

The variational parameters $\theta$ are optimized by minimizing the expectation value of the Hamiltonian,
\[
\mathcal{L}_{\theta}
=
\frac{\langle \psi_{\theta} \mid \hat{H} \mid \psi_{\theta} \rangle}
{\langle \psi_{\theta} \mid \psi_{\theta} \rangle},
\]
which serves as the loss function during the training process. 
We use the Adam optimizer with moments $\beta_{1}=0.9$ and $\beta_{2}=0.999$. 
In the initial phase of optimization, we first train the phase part of the network while keeping the amplitude part fixed, 
as it facilitates the network to learn the wavefunction sign structure
more effectively \cite{Szab__2020}. 

For the $\rho$-parameters, the learning rate schedule begins with a linear warm-up from 0 to 0.001 during the first 5,800 iterations, 
remains constant at 0.001 up to 16,000 iterations, and then reduced to 0.0001 for the remaining iterations. 
For the $\phi$-parameters, the learning rate is maintained at 0.001 until 16,000 iterations and then decreased to 0.0001 for the remaining iterations.
The entire training process runs for 19,000 iterations. 
Sampling over spin configurations is carried out using the Markov Chain Monte Carlo technique. 
We use $10^{4}$ Monte Carlo samples to estimate the energy expectation value during training, 
whereas $10^{7}$ samples are generated for evaluating other observables after the training \cite{Joshi2023GroundSP, kumar2025impactrandombonddisorder}. 
The complete implementation of the neural network architecture and the sampling procedure is performed using the 
NetKet library \cite{netket3:2022}.
To compute local spin expectation values, we use $\langle S_i \rangle = \langle \sigma_i /2 \rangle$, where $\sigma_i$ is the NetKet spin operator. 

\textit{Classical Monte Carlo Simulations}---For the classical simulations, we employ the Monte Carlo method based on the Metropolis--Hastings algorithm. 
The first 40,000 Monte Carlo steps are used for annealing, during which the temperature is gradually reduced 
from $T = 1 J_{1}$ to $T = 0.01 J_{1}$. 
This is followed by additional 40,000 steps for thermalization. 
Finally, 60,000 Monte Carlo steps are performed for measuring physical observables, 
with data sampling executed after every 200 steps to minimize autocorrelation effects.

\bibliographystyle{apsrev4-2}
\bibliography{references}

\end{document}